\documentclass[conference]{IEEEtran}
\IEEEoverridecommandlockouts
\usepackage{cite}
\usepackage{amsmath,amssymb,amsfonts}
\usepackage{algorithmic}
\usepackage{graphicx}
\usepackage{textcomp}
\usepackage{xcolor}

\usepackage{multirow}
\usepackage{makecell}
\usepackage{pbox}
\usepackage{breakurl}
\usepackage{url}
\usepackage{todonotes}

\def\BibTeX{{\rm B\kern-.05em{\sc i\kern-.025em b}\kern-.08em
    T\kern-.1667em\lower.7ex\hbox{E}\kern-.125emX}}
        
\usepackage{fancyhdr}
\begin{document}

\title{Engineering Music to Slow Breathing and Invite Relaxed Physiology}

\author{\IEEEauthorblockN{Grace Leslie$^*$\thanks{$^*$Equal contribution}}
\IEEEauthorblockA{\textit{School of Music, Georgia Tech}\\
Atlanta, GA, US \\
grace.leslie@gatech.edu}
\and
\IEEEauthorblockN{Asma Ghandeharioun$^*$}
\IEEEauthorblockA{\textit{MIT Media Lab}\\
Cambridge, MA, US \\
asma\_gh@mit.edu}
\and
\IEEEauthorblockN{Diane Zhou}
\IEEEauthorblockA{\textit{MIT}\\
Cambridge, MA, US \\
dianez@mit.edu}
\and
\IEEEauthorblockN{Rosalind W. Picard}
\IEEEauthorblockA{\textit{MIT Media Lab}\\
Cambridge, MA, US\\
picard@media.mit.edu}
}

\maketitle

\begin{abstract}
We engineered an interactive music system that influences a user's breathing rate to induce a relaxation response. This system generates ambient music containing periodic shifts in loudness that are determined by the user's own breathing patterns. We evaluated the efficacy of this music intervention for participants who were engaged in an attention-demanding task, and thus explicitly not focusing on their breathing or on listening to the music. We measured breathing patterns in addition to multiple peripheral and cortical indicators of physiological arousal while users experienced three different interaction designs: (1) a  ``Fixed Tempo" amplitude modulation rate at six beats per minute; (2) a ``Personalized Tempo" modulation rate fixed at  75\% of each individual's breathing rate baseline, and (3) a ``Personalized Envelope" design in which the amplitude modulation matches each individual's breathing pattern in real-time. Our results revealed that each interactive music design slowed down breathing rates, with the ``Personalized Tempo" design having the largest effect, one that was more significant than the non-personalized design. The physiological arousal indicators (electrodermal activity, heart rate, and slow cortical potentials measured in EEG) showed concomitant reductions, suggesting that slowing users' breathing rates shifted them towards a more calmed state. These results suggest that interactive music incorporating biometric data may have greater effects on physiology than traditional recorded music.
\end{abstract}

\begin{IEEEkeywords}
music, intervention, breathing, EEG, ECG, EDA, relaxation, stress.
\end{IEEEkeywords}

\section{Introduction}

Music invites strong emotional responses in its listeners, and thus presents a promising avenue for the design of interaction systems that bring health and well-being to users. Historically, researchers in the music cognition community have attributed musical emotion to cognitive appraisal \cite{meyer1956emotion}; however, more recent models acknowledge the possibility of multiple mechanisms by which music evokes emotions \cite{van2012directing, dijk2011breathe}, including brain stem reflexes and emotional contagion \cite{juslin2008emotional}. In fact, listeners have reported many different physical reactions due to focused music listening \cite{sloboda1991music}. Music has been shown to help alleviate pain \cite{newbold2015musically}, influence movement synchronization \cite{aschersleben2002temporal, repp2005sensorimotor}, and be used as a bio-feedback signal \cite{bergstrom2014using}. In the present research, we explore the possibility that indirect exposure to musical stimuli may also affect users in such a way as to produce beneficial changes in their physiology, without requiring attentive listening.

\enlargethispage{1\baselineskip}

Past research has detailed the range of both perceived and felt emotions that accompany attentive music listening \cite{zentner2008emotions}, in addition to a range of physical \cite{jaimovich2012emotion} and physiological \cite{Sammler2007}\cite{Salimpoor2009} responses associated with emotions. However, less is known about how music may be specifically designed to induce a particular physiological response in the listener. Previous studies have shown that influencing breathing may be one effective pathway to inviting shifts in physiology. For example, instructing users with music in the control of their breathing patterns showed promise in breath regulation \cite{siwiak2009catch}, relaxation \cite{yu2018unwind}, creating a meditative experience \cite{vidyarthi2014interactively}, and reducing muscle tension \cite{robb2000music}. However, such studies relied on intentional breathing to be effective. Other studies have shown that listening to music can effect breathing rates \cite{bernardi2006cardiovascular}, introducing white noise to music can influence electrodermal activity and heart rate variability \cite{bhandari2015music}, though these were not carefully controlled to avoid intentional manipulation of breathing. The present study was carefully designed to examine any effects that auditory feedback may have on breathing even when the listener is unaware that an intervention is taking place, remaining fully engaged in a demanding task.

Breathing is a promising avenue to explore for regulation of affective state, as it is a physiological process that is under autonomic control, yet which can also be controlled externally through conscious effort or external influences \cite{moraveji2011peripheral}. The rate and manner of one's breathing pattern changes with physical exertion \cite{robertson1982central} and affective state \cite{charmandari2005endocrinology, bloch1991specific}; in turn, aerobic capacity, stress level \cite{brown2005sudarshana, brown2005sudarshanb}, mental functions \cite{soni2015effect}, and mood \cite{philippot2002respiratory} can be influenced by consciously manipulating breathing, as is detailed in ancient Yoga Sutras \cite{bryant2015yoga}, among more modern scientific research.

It is known that reducing one's breathing rate can reduce perceived momentary stress levels \cite{brown2005sudarshana, brown2005sudarshanb}. While consciously slowing down breathing using mindfulness exercises can be effective \cite{kabat2003mindfulness}, such interventions require focused attention and divert attention from important tasks \cite{adams2015mindless}, which can make them impractical in workplaces. Researchers are beginning to explore implicit bio-feedback\cite{kuikkaniemi2010influence}. Recent findings showed that rhythmically oscillating audiovisual feedback presented to users significantly influenced slower breathing, had a lasting effect, improved self-reported calmness and focus, and was highly preferred for future use \cite{ghandeharioun2017brightbeat}. To this end, we designed three interactive music systems incorporating rhythmic loudness changes, also using a breath sensor (Zephyr BioHarness), to probe which interactions optimally, yet effortlessly, influence breathing rate through sound. 

\enlargethispage{1\baselineskip}

\section{Theory and Design}
\begin{figure*}[!t]
  \includegraphics[width=1.0\columnwidth]{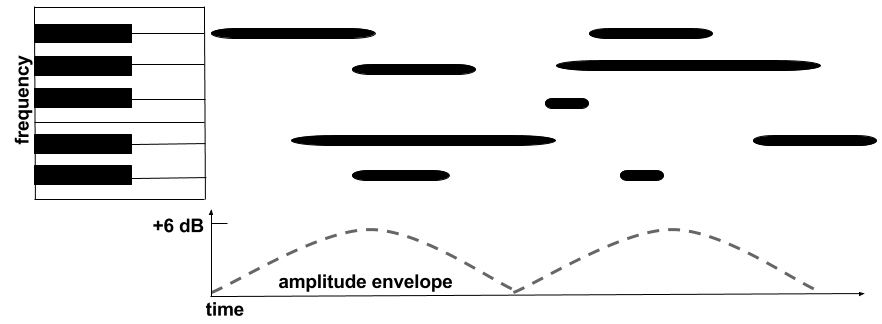}
  \hspace{1cm}
  \includegraphics[width=0.9\columnwidth]{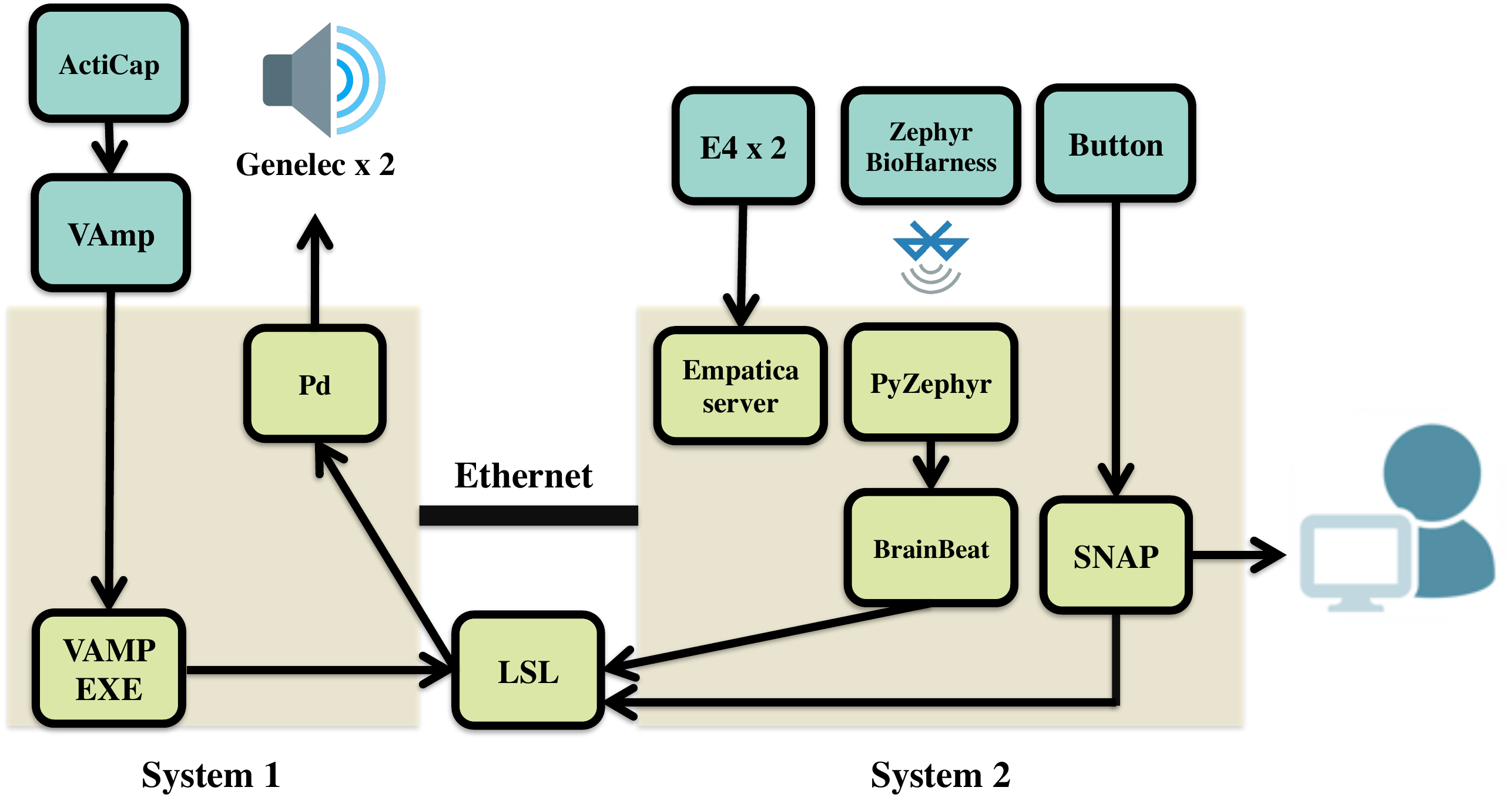}
  \caption{System design. Left: The loudness of the music was modulated to produce an undulating, breath-like sound. Right: Top-level system view. Color codes: blue: hardware module, green: software module.}~\label{fig:system}
\end{figure*}

This interactive music system was designed to influence the user's breathing pattern in order to invite a relaxation response. It was engineered with two competing principles in mind: first, this system must be as unobtrusive as possible, so as to not require focused attention that would detract from everyday activities or workplace tasks. We also wanted to avoid any prior familiarity with the music that might have prior emotional associations.
Second, this system must invite physiological changes in the listener that accompany a relaxed state.  
To satisfy the first constraints, we engineered an original ambient-genre music composition using the PureData (Pd)\cite{puckette1996pure} graphical programming language. Our software took an incoming stream of real-time breathing data recorded from the participant's Zephyr BioHarness (https://www.zephyranywhere.com) and translated this data to control the loudness of the principal melodic line in the ambient music mix. An amplitude envelope, calculated using a square-root function normally used to generate stereo panning curves, was applied to the overall sound mixture in order to produce an undulating volume effect reminiscent of the time course of a normal inhale and exhale breathing cycle. A diagram of this envelope is included in the design illustration in Fig. \ref{fig:system}. The difference in overall loudness between full ``inhale" maxima and full ``exhale" minima was 6 dB, representing a doubling of loudness during the course of a synthesized breath cycle. A sample of this music can be streamed at \url{http://soundcloud.com/gracesounds}.

We developed three intervention designs that differed only in the level of interaction between the musical composition engine and the participant's real-time breathing data stream:

\subsection{Fixed Tempo (FT)}
In the Fixed Tempo design, the synthesized breathing cycle was set at 6 breaths per minute (bpm), a rate shown to produce ideal levels of relaxation \cite{lehrer2000resonant}. Therefore, this condition was equivalent to playing a pre-recorded  piece of music designed to have relaxing qualities, but lacking individualization to each user's breathing pattern.

\subsection{Personalized Tempo (PT)}
We calculated each participant's average baseline breathing rate at the beginning of the experiment session. During the Personalized Tempo block, the music was presented with amplitude modulations occurring at 75\% of the calculated baseline breathing rate, in an effort to gently ``nudge" a reduction in breathing rate. In any case of a participant having an abnormally high natural breathing rate, their Personalized Tempo was capped at 15 beats per minute.

\subsection{Personalized Envelope (PE)} 
In the most interactive design, the real-time breathing data stream fully governed the rate of synthesized ``inhale" and ``exhale," producing a sympathetic music feedback which mirrored the exact time course of each participant's unintentional inhales and exhales. While PT relies on a single number calculated during a participant's Baseline session, PE uses one's instantaneous breathing signal to generate each note.

\section{Methods}

\subsection{Experimental Procedure}
This study was pre-approved by the MIT Review Board. Our sample featured 19 participants (11 females, 8 males), of varying ages (19--55 years). 

The experiments were conducted indoors in a sound-treated studio with no audible external noise. Participants were seated at a table in front of a laptop at a distance of approximately 2 meters from each of a pair of studio audio monitors (Genelec) placed at approximately 30\textdegree  left and right of center. During data collection, participants were asked to keep still, breathe spontaneously, and fixate their vision on a computer-generated crosshair symbol. We also recorded their breathing waveform, and electrocardiogram (ECG) data using a Zephyr BioHarness (Zephyr\textsuperscript{TM} Performance Systems) sampled at 17 Hz, and 250Hz, respectively. Bilateral electrodermal activity (EDA) at 4 Hz was recorded from right and left wrists using the E4 wristband (Empatica, Inc). See the system diagram in Fig. \ref{fig:system}.

Prior to data collection, the wristbands and Zephyr Bioharness were placed on each participant, and they were asked to walk up and down three flights of stairs in order to properly prime the EDA, ECG, and respiration sensors. After the sensor priming, participants were outfitted with a 16-channel EasyCap EEG cap connected to a BrainVision VAmp EEG amplifier. 

Participants completed four blocks of 40 forewarned reaction time trials; each block lasted an average of seven minutes due to random inter-trial intervals of between two and five seconds. For each trial, a warning stimulus was presented, followed by a fixed silent interval of 4.5 seconds. After this fixed interval, the imperative stimulus, an alarming short buzzing sound, was played. The participants were instructed to press a key on a computer keyboard as soon as they hear the imperative stimulus. This trial design is a classic stimulus presentation from the event-related potential (ERP) literature in EEG research, and is designed to elicit the contingent negative variation (CNV), a slow cortical potential whose amplitude increases with the focusing of attentional resources required for this task\cite{Tecce1972}. Since this task is demanding, it provides the opportunity to test the calming influence of our interventions on peripheral and cortical arousal.

In order to minimize muscle artifacts in EEG, participants were reminded not to blink or move their eyes unless absolutely necessary, to keep their eyes open as much as possible while still blinking naturally, to keep their eyes fixated on the displayed crosshair, and to not clench their muscles or move their right hand unnecessarily, even when they were trying to respond very quickly to the buzzer sound.

The first of the four blocks was presented silently, in order to collect baseline physiological measurements. In each of the subsequent three blocks, one of the interactive music interventions was introduced at random. 

The Lab Streaming Layer \cite{kothe2014lab} protocol provided real-time synchronization of stimulus presentation timestamps along with breathing, ECG, and EEG streams. EDA data were synchronized with the other streams post-hoc using the timestamps recorded on the E4. 

All participants provided prior informed consent for the primary reaction time task with accompanying musical stimuli. Participants were told that the music was provided as entertainment to mitigate possible boredom during long blocks of the repetitive task, and that multiple physiological measurements would be taken in order to test the effects of focused attention on these bodily processes. This deceptive experiment design was necessary to ensure that participants would not attempt to consciously entrain their breathing to the music, or otherwise manipulate their breathing rate to achieve any particular result. After data collection was complete, each participant was informed of the real purpose of the experiment.

\subsection{Pre-processing Steps}
\label{sec:preprocessing}

The raw breathing waveform was measured as a time-series representation of the extent to which the participants' breathing increased and decreased tension in the BioHarness chest strap. The breathing waveform data were filtered using a lowpass Butterworth filter with a cutoff frequency of 1 Hz, representing the fastest reasonable breathing rate. Local maxima of the breathing signal were calculated. We enforced the detected peaks to drop at least 2 $nu$s\footnote{$nu$ refers to non-unit, a relative pressure unit for raw breathing waveform.} on either side before the signal attains a higher value. After detecting the peak locations, the inter-respiration intervals (IRI) were calculated in milliseconds. To better highlight the influence of interventions and lower the influence of personal baselines, we standardized measures for each participant. Specifically, the inter-respiration intervals were z-scored per participant using the following formula:
\begin{equation}
    z_{IRI} = \frac{IRI-\mu_{IRI}}{\sigma_{IRI}}
    \label{z-score}
\end{equation}
Here, $z_{IRI}$ refers to the z-score of IRIs. Moreover, $\mu_{IRI}$ refers to the mean of the signal during a complete session of the participant, including Baseline, Fixed Tempo, Personalized Envelope, and Personalized Tempo conditions. Similarly, $\sigma_{IRI}$ refers to the standard deviation of the IRIs during the whole session. The mean of $z_{IRI}$ was calculated per block per participant as a proxy for the inverse of the relative breathing rate. The standard deviation of $z_{IRI}$ was calculated per block per participant as a proxy for variability of breathing.

Raw EDA was measured using the E4 sensors in $\mu S$. We applied a 6th order Butterworth low-pass filter (1 Hz cutoff frequency) to the EDA data. The filtered EDA was transformed into z-scores ($z_{EDA}$) similar to the method described in Equation \ref{z-score}. To better capture the relaxation response, we focused on the tonic skin conductance level which measures the smooth underlying slowly changing levels of EDA. Specifically, we measured the rate of change in the skin conductance level per experiment block per participant using the following formula:
\begin{equation}
    slope_{z_{EDA}} = fs \times (z_{EDA[2:n]}-z_{EDA[1:n-1]})
\end{equation}
Here, $fs$ refers to the frequency of sampling which is 4Hz. Also, $n$ refers to the length of the EDA signal.

All EEG data processing was performed offline using the EEGLAB \cite{delorme2004eeglab} toolbox for Matlab. Continuous EEG data were high-pass filtered at .05 Hz by subtracting a low-passed version of the signal (.05 Hz) from the original signal. Channels with a kurtosis greater than five were excluded from analysis, and the remaining channels were re-referenced to their average. The continuous EEG data were epoched around the first audio stimulus of each trial, from 1 second prior to the stimulus to 4 seconds post-stimulus, and the 500ms preceding the stimulus was used as the baseline. Individual epochs with absolute amplitudes exceeding 50uV were excluded from analysis. We calculated the mean amplitude in the Cz channel over three time windows representing the early (400-1400 ms), mid (1500-2600 ms), and late-stage (2600-3700 ms) CNV \cite{Funderud2012}. 

We used the Pan-Tompkins implementation of the QRS complex detector for ECG analyses \cite{pan1985real}. The R-R intervals, i.e. inter-beat intervals (IBI), were consequently calculated. We used the Python HRV package \cite{hrvpackage} to calculate a range of time-based, frequency-based, and non-linear heart rate variability features from the IBIs. We also transformed the IBIs into z-scores similarly as in Equation (\ref{z-score}) by standardizing IBIs within each session ($z_{IBI}$).

\section{Results and Discussion}

\subsection{Comparing Interventions}
\label{sec:interventions}

In this section, we present our findings regarding physiological changes arising from ambient music conditions in comparison to baseline. We use box-plot \cite{frigge1989some}, where the middle line represents median, the inner-box covers first to third quartiles, and the whiskers extend the inner-box boundaries 1.5 times inter-quartile range. 

\subsubsection{Breathing}
\label{sec:breathing}

\begin{figure}[!t]
  \centering
  \includegraphics[width=0.9\columnwidth]{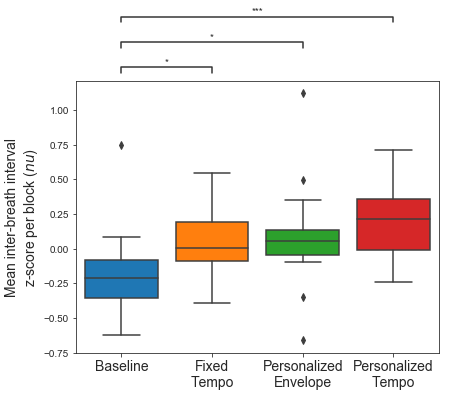}
  \caption{Comparison of the average of inter-respiration interval z-scores across conditions. Higher inter-breath intervals means lower relative breathing rates, and is associated with more relaxed states. See \S\ref{sec:interventions} for box-plot details \& \S\ref{sec:breathing} for ANOVA statistics. Squared brackets show p-values of the post-hoc independent t-tests. **:$.001<p \le .01$, ***: $.0001<p \le .001$.}
  \label{fig:iri-mean}
\end{figure}

Relaxation has multiple physiological indicators, including changes in respiration patterns. Deep and slow breathing both arise from and give rise to physiological, affective, and cognitive calm \cite{grossman2001breathing}. On the other hand, sustained attention and cognitive load have been shown to reduce respiratory variability \cite{vlemincx2012sigh}. Moreover, negative emotional states are shown to reduce correlated breathing variability \cite{vlemincx2011sigh} and sense of relief is associated with higher breathing variability \cite{vlemincx2010take}. In the aforementioned articles, tidal volume, instantaneous respiration rate, and minute ventilation and their coefficient of variation have been used to measure total respiratory variability. Additionally, autocorrelation at one breath lag has been used to quantify correlated respiratory variability. Breathing variability is also a predictor of respiratory health and has been associated with more successful separation of the patient from the ventilator \cite{wysocki2006reduced, baudin2014impact}. Given the rich literature on breathing and its relationship with affective states and wellbeing, there is a consensus that slower breathing rate and higher breathing variability are associated with a calmer state.

To quantify effects on calming, we thus focus on two metrics from the breathing signal: the z-score of inter-respiration intervals within each experiment block ($z_{IRI}$), as well as the variance of inter-respiration interval z-scores for each block ($var_{z_{IRI}}$). See \S\ref{sec:preprocessing} for more information about how to calculate these features. $z_{IRI}$ is proportional to the inverse of relative breathing rate. Thus we expect to see higher $z_{IRI}$ in a more relaxed state. Additionally, $var_{z_{IRI}}$ is associated with breathing variability. Therefore, we expect to see higher $var_{z_{IRI}}$ in more positive and calming settings.

As shown in Fig. \ref{fig:iri-mean}, our analyses revealed that there was a significant difference in z-score of inter-respiration intervals ($z_{IRI}$) across all the conditions. An ANOVA test was performed to evaluate differences in $z_{IRI}$ between Baseline, Fixed Tempo, Personalized Envelope, and Personalized Tempo designs: 
$F(3,72)=5.228, p=.003$. 
We conducted post-hoc pairwise comparisons using the independent t-test to compare each music design condition to the baseline: 
$t_{B-FT}=87.0, p_{B-FT}=.007;$
$t_{B-PE}=74.0, p_{B-PE}=.002;$
$t_{B-PT}=48.0, p_{B-PT}=.000;$
The Personalized Tempo design increases $z_{IRI}$ the most. The Personal Envelope design may influence breathing most strongly, but because of the variance between fast and slow breathers, it has a lesser effect overall. Moreover, simply having a fixed slow music tempo also reduces breathing (Fixed Tempo design), but has a lesser effect than the personalized designs. 

\begin{figure}[!t]
  \centering
  \includegraphics[width=0.85\columnwidth]{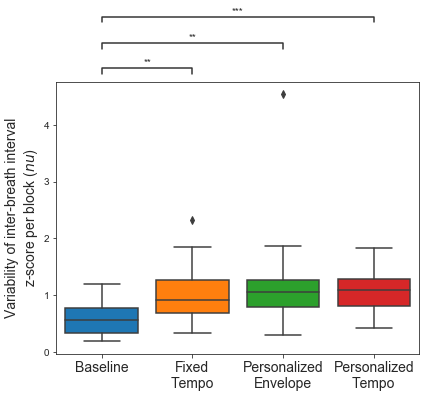}
  \caption{Comparison of the variability of inter-respiration interval z-scores across conditions. Higher variability in breathing is associated with more relaxed states. See \S\ref{sec:interventions} for box-plot details \& \S\ref{sec:breathing} for ANOVA statistics. Squared brackets show p-values of the post-hoc independent t-tests. **:$.001<p \le .01$, ***: $.0001<p \le .001$.}
  \label{fig:iri-var}
\end{figure}

Additionally, our analyses revealed that there was a significant difference in variability of inter-respiration interval z-scores ($var_{z_{IRI}}$) across all the conditions, as shown in Fig. \ref{fig:iri-var}. An ANOVA test was performed to evaluate differences in $var_{z_{IRI}}$ between Baseline, Fixed Tempo, Personalized Envelope, and Personalized Tempo designs: 
$F(3,72)=3.962, p=.011$. 
We conducted post-hoc pairwise comparisons using the independent t-test to compare each music design condition to the baseline: 
$t_{B-FT}=88.0, p_{B-FT}=.007;$
$t_{B-PE}=66.0, p_{B-PE}=.001;$
$t_{B-PT}=65.0, p_{B-PT}=.001;$
We observe that the music conditions resulted in a more variable breathing pattern which is associated with a more positive state. Moreover, the difference in breathing variability is more prominent in the personalized designs, suggesting that there is added value in bringing physiology-driven design to ambient music listening.

\subsubsection{Electrodermal Activity (EDA)}
\label{sec:eda}

\begin{figure}[!t]
  \centering
  \includegraphics[width=0.9\columnwidth]{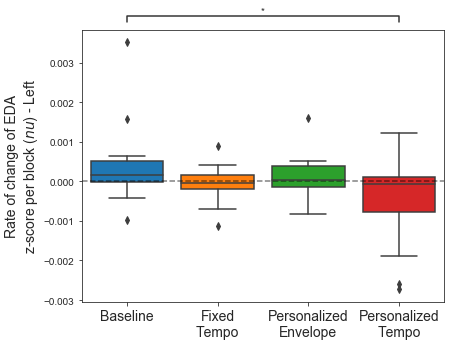}
  \caption{Comparison of the rate of change of tonic EDA z-score across conditions. Negative values mean decreasing EDA levels, and are associated with a relaxing effect. See \S\ref{sec:interventions} for box-plot details \& \S\ref{sec:eda} for ANOVA statistics. Squared brackets show p-values of the post-hoc independent t-tests. *:$.01<p \le .05$.}
  \label{fig:eda}
\end{figure}
EDA is traditionally characterized into two types, tonic and phasic activity. Tonic activity or skin conductance level shows the slowly-changing patterns of EDA. Lower levels of tonic activity are associated with more calming states. However, phasic activity or skin conductance response corresponds to rapid changes in EDA level in the form of peaks with a particular morphology: a quick rise and a slow decay. Higher skin conductance is usually associated with higher sympathetic stimulation and higher stress \cite{boucsein2012electrodermal}. Since our experiment design presented stimuli with the goal of creating a relaxation response, we expect the tonic portion of the EDA to be more indicative of the effectiveness of our interactive music. As recommended by \cite{boucsein2012electrodermal}, we use z-scoring for standardization of raw EDA and use $slope_{z_{EDA}}$ as a metric of rate of change in skin conductance level. See \S\ref{sec:preprocessing} for more information about feature calculation.
In our study, participants were using their dominant hand for doing the primary reaction-time task and making mouse clicks continuously. This resulted in motion artifacts introduced to the right hand EDA signal. Thus, for this analysis, we focus on the non-dominant EDA, which has a long history of study in the skin conductance literature \cite{boucsein2012electrodermal}.

As shown in Fig. \ref{fig:eda}, our analyses revealed that there was a significant difference in rate of change in skin conductance level ($slope_{z_{EDA}}$) across all the conditions. An ANOVA test was performed to evaluate differences in $slope_{z_{EDA}}$ between Baseline, Fixed Tempo, Personalized Envelope, and Personalized Tempo designs: 
$F(3,68)=3.158, p=.030$\footnote{The left E4 sensor did not record any data for one participant. This resulted in 4 missing data-points, one per condition, for this analysis.}. We conducted post-hoc pairwise comparisons using the independent t-test to compare each music condition to the baseline. We only observed significant differences across the Personalized Tempo design and the baseline\footnote{$t_{B-FT}=219.0, p_{B-FT}=.074;$ $t_{B-PE}=191.0, p_{B-PE}=.367;$
}:
$t_{B-PT}=228.0, p_{B-PT}=.038;$ 
This analysis shows that the Personalized Tempo design was specifically more influential in inducing a physiological state of calm and resulted in decreasing levels of tonic EDA.
 
\subsubsection{Electroencephalogram (EEG)}
\label{sec:eeg}

\begin{figure}[!t]
  \centering
  \includegraphics[width=0.9\columnwidth]{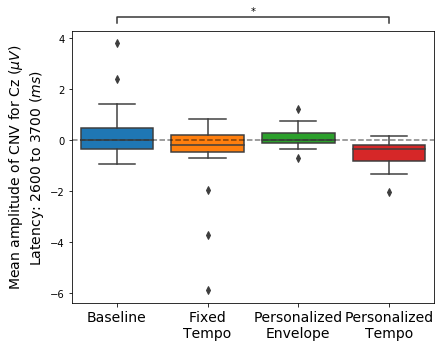}
  \caption{Comparison of the mean amplitude of contingent negative variation (CNV) for Cz electrode across conditions. Lower CNV with greater absolute amplitude shows higher cortical arousal. In line with previous research findings\cite{LIM1996151, higuchi1997effects}, cortical and peripheral arousal have an inverse relationship. See \S\ref{sec:interventions} for box-plot details \& \S\ref{sec:eeg} for ANOVA statistics. Squared brackets show p-values of the post-hoc independent t-tests. *:$.01<p \le .05$.}
  \label{fig:eeg}
\end{figure}

We gave participants a reaction-time task that was chosen to elicit the contingent negative variation, a well-studied slow cortical potential that is known to index cortical arousal \cite{Tecce1972}. Past studies have shown that a decrease in EDA due to habituation to an auditory stimulus is accompanied by an increase in measured EEG power\cite{LIM1996151, higuchi1997effects}. Others have demonstrated that relaxation-inducing biofeedback causing a decrease in autonomic activity measured peripherally was also accompanied by an increase in the CNV amplitude \cite{nagai2004influence}.
As shown in Fig. \ref{fig:eeg}, our analyses revealed that there was a significant difference between the amplitude of the late-stage CNV when compared between the baseline condition and all three design conditions. An ANOVA test was performed to evaluate differences in the late-stage CNV amplitude between the Baseline, Fixed Tempo, Personalized Envelope, and Personalized Tempo designs: 
$F(3,61)=3.180, p=.031$\footnote{Note that 8 datapoints had kurtosis greater than 5 (2 per each condition) and 3 were detected as outliers for having a z-score of more than 3 or less than -3 (1 per each intervention condition).}. 
We conducted post-hoc pairwise comparisons using the independent t-test to compare each music design condition to the baseline. We only observed significant differences across the Personalized Tempo design and the baseline
\footnote{$t_{B-FT}=1.988, p_{B-FT}=.056;$
$t_{B-PE}=0.600, p_{B-PE}=.553;$ 
}: 
$t_{B-PT}=2.556, p_{B-PT}=.016;$ 
The analysis of our breathing and EDA data shows that the Personalized Tempo design was specifically more influential in inducing a physiological state of calm. The accompanying result showing an influence on slow cortical potentials suggests that the decrease in peripherally measured autonomic activity caused an increase in cortical excitability, replicating findings of previous independent studies \cite{LIM1996151, higuchi1997effects}.

\subsubsection{Electrocardiogram (ECG)}
\label{sec:ecg}

\begin{figure}[!t]
  \centering
  \includegraphics[width=0.85\columnwidth]{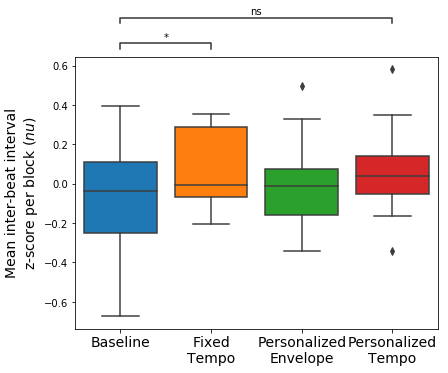}
  \caption{Comparison of the average of inter-beat interval (IBI) z-scores across conditions. Higher IBI is associated with lower heart rate, usually accompanying a relaxed state. See \S\ref{sec:interventions} for box-plot details \& \S\ref{sec:ecg} for ANOVA statistics. Squared brackets show p-values of the post-hoc independent t-tests. ns: $.05<p \le .1$, *:$.01<p \le .05$.}
  \label{fig:ibi-mean}
  \vspace{-.5cm}
\end{figure}

We calculated the z-score of the inter-beat intervals ($z_{IBI}$). See \S\ref{sec:preprocessing} for more information about this feature. $z_{IBI}$ is proportional to the inverse of relative heart rate; thus we expect to see a higher $z_{IBI}$ in a more relaxed state. 
Fig. \ref{fig:ibi-mean}, visualizes the difference between inter-beat interval z-scores ($z_{IBI}$) across all conditions. An ANOVA test was performed to evaluate differences in $z_{IBI}$ between Baseline, Fixed Tempo, Personalized Envelope, and Personalized Tempo designs:
$F(3, 68)=1.984, p=.125$\footnote{Experiment blocks with less than two minutes of valid IBIs (one data point per each condition) were excluded from this analysis.}.
We also report pairwise comparisons using an independent t-test that compare each music design condition to the baseline. We only observed a significant difference between Fixed Tempo and Baseline conditions\footnote{$t_{B-PE}=-1.159, p_{B-PE}=.255;$ $t_{B-PT}=-1.827, p_{B-PT}=.077;$ 
}:
$t_{B-FT}=-2.104, p_{B-FT}=.043;$ 

Though we calculated a comprehensive list of heart rate variability (HRV) features from ECG \cite{hrvpackage}, we did not observe any significant differences between baseline and intervention conditions. 
This finding is to be expected, given that traditional HRV features may not best represent how the autonomic nervous system (ANS) is influencing heart function \cite{aysin2006effect}. HRV is controlled by both the sympathetic and parasympathetic branches of the ANS. Traditional HRV measures are incapable of isolating the effects of these two branches, especially while breathing is changing. Particularly at low respiration rates, the parasympathetic activity shifts into lower frequencies and overlaps with the frequency interval that is traditionally associated with sympathetic activity \cite{aysin2006effect}. Given that our ambient music design conditions have resulted in slower and more variable breathing, traditional HRV features are not equipped to distinguish between parasympathetic and sympathetic control of the heart \cite{aysin2006effect}. See \S\ref{sec:limitations} for future work directions to try to mitigate this problem.

\subsection{Limitations}
\label{sec:limitations}
In this paper, we solely focused on the left wrist EDA signal due to the motion artifacts introduced to the EDA captured from the right wrist. Recent findings show that a more holistic view of the EDA signal from multiple locations on the body could reveal further information about the affective and cognitive state of the user \cite{picard2016multiple}. In the future, we would like to explore alternative signal processing techniques to overcome motion artifacts and study EDA asymmetry features.

We focused on traditional HRV measures for studying the influence of ANS on the heart. However, we learned that these features are not suitable over slow and variable breathing rates \cite{aysin2006effect}. For future work, we would like to explore enhanced HRV features \cite{aysin2006effect} that better distinguish sympathetic and parasympathetic indices of HRV in slow and variable breathing.

In addition, we did not inquire of participants after the experiment if they suspected that the real intention of the study was to manipulate their breathing pattern. In the future, these steps will be taken to ensure that analyzed data is only from participants unaware of the the music presentation intention. Future experiments are needed to test generalizability to real-world tasks and test if the influence of the music intervention extends to arousing effects using uptempo, arousing stimuli.

\section{Conclusion}
This study engineered music in a systematic way to influence breathing in participants, despite that their focus was on a (different) cognitive task. The three music interactions differed with respect to the amount of customization to the participant's own natural breathing rates. While previous designs have shown promise in encouraging entrainment of breathing patterns to external stimuli, this study was the first of its kind to specifically target unfocused entrainment using a deceptive experiment design to ensure participants did not consciously mimic the stimuli with their breath. 

Our results revealed that the intricate design of musical stimuli did influence the participants' breathing patterns. All intervention conditions resulted in higher relative inter-breath intervals, i.e. lower relative breathing rates, which are associated with a calmer state. Additionally, they influenced the relative variability of breathing which is also associated with a more positive and relaxed state. Importantly, personalizing the system based on the user's natural breathing rate made the system significantly more influential to the user's breathing pattern in addition to measures of their relaxation response, as seen in the downward shift in the tonic activity of EDA. Similarly, it resulted in greater cortical arousal as measured by the CNV, which has been shown to have an inverse association with peripheral arousal.

\section{Acknowledgments}
We thank Brain Vision LLC (Cary, NC, USA) for providing the EEG equipment and MIT Media Lab Consortium for supporting this research.

\bibliographystyle{IEEEtran}
\bibliography{references}

\begin{thebibliography}{10}
\providecommand{\url}[1]{#1}
\csname url@samestyle\endcsname
\providecommand{\newblock}{\relax}
\providecommand{\bibinfo}[2]{#2}
\providecommand{\BIBentrySTDinterwordspacing}{\spaceskip=0pt\relax}
\providecommand{\BIBentryALTinterwordstretchfactor}{4}
\providecommand{\BIBentryALTinterwordspacing}{\spaceskip=\fontdimen2\font plus
\BIBentryALTinterwordstretchfactor\fontdimen3\font minus
  \fontdimen4\font\relax}
\providecommand{\BIBforeignlanguage}[2]{{%
\expandafter\ifx\csname l@#1\endcsname\relax
\typeout{** WARNING: IEEEtran.bst: No hyphenation pattern has been}%
\typeout{** loaded for the language `#1'. Using the pattern for}%
\typeout{** the default language instead.}%
\else
\language=\csname l@#1\endcsname
\fi
#2}}
\providecommand{\BIBdecl}{\relax}
\BIBdecl

\bibitem{meyer1956emotion}
L.~B. Meyer, \emph{Emotion and meaning in music}.\hskip 1em plus 0.5em minus
  0.4em\relax U. of Chicago P., 1956.

\bibitem{van2012directing}
M.~D. van~der Zwaag, J.~H. Janssen, and J.~H. Westerink, ``Directing physiology
  and mood through music: Validation of an affective music player,''
  \emph{TAC}, vol.~4, no.~1, pp. 57--68, 2012.

\bibitem{dijk2011breathe}
E.~O. Dijk and A.~Weffers, ``Breathe with the ocean: a system for relaxation
  using audio, haptic and visual stimuli,'' in \emph{EuroHaptics}, 2011.

\bibitem{juslin2008emotional}
P.~N. Juslin and D.~V{\"a}stfj{\"a}ll, ``Emotional responses to music: The need
  to consider underlying mechanisms,'' \emph{Behav. and brain sciences},
  vol.~31, no.~5, pp. 559--575, 2008.

\bibitem{sloboda1991music}
J.~A. Sloboda, ``Music structure and emotional response: Some empirical
  findings,'' \emph{Psychology of music}, vol.~19, no.~2, pp. 110--120, 1991.

\bibitem{newbold2015musically}
J.~Newbold, N.~Berthouze, N.~Gold, and A.~Williams, ``Musically informed
  sonification for self-directed chronic pain physical rehabilitation,'' 2015.

\bibitem{aschersleben2002temporal}
G.~Aschersleben, ``Temporal control of movements in sensorimotor
  synchronization,'' \emph{Brain and cognition}, vol.~48, no.~1, pp. 66--79,
  2002.

\bibitem{repp2005sensorimotor}
B.~H. Repp, ``Sensorimotor synchronization: a review of the tapping
  literature,'' \emph{Psychonomic bul. \& rev.}, vol.~12, no.~6, 2005.

\bibitem{bergstrom2014using}
I.~Bergstrom, S.~Seinfeld, J.~Arroyo-Palacios, M.~Slater, and M.~V.
  Sanchez-Vives, ``Using music as a signal for biofeedback,''
  \emph{International J. of Psychophysiology}, vol.~93, no.~1, pp. 140--149,
  2014.

\bibitem{zentner2008emotions}
M.~Zentner, D.~Grandjean, and K.~R. Scherer, ``Emotions evoked by the sound of
  music: characterization, classification, and measurement.'' \emph{Emotion},
  vol.~8, no.~4, p. 494, 2008.

\bibitem{jaimovich2012emotion}
J.~Jaimovich, N.~Coghlan, and R.~B. Knapp, ``Emotion in motion: A study of
  music and affective response,'' in \emph{International Symposium on Computer
  Music Modeling and Retrieval}.\hskip 1em plus 0.5em minus 0.4em\relax
  Springer, 2012, pp. 19--43.

\bibitem{Sammler2007}
D.~Sammler, M.~Grigutsch, T.~Fritz, and S.~Koelsch, ``{Music and emotion:
  electrophysiological correlates of the processing of pleasant and unpleasant
  music.}'' \emph{Psychophysiology}, vol.~44, no.~2, 2007.

\bibitem{Salimpoor2009}
V.~N. Salimpoor, M.~Benovoy, G.~Longo, J.~R. Cooperstock, and R.~J. Zatorre,
  ``{The rewarding aspects of music listening are related to degree of
  emotional arousal.}'' \emph{PloS one}, vol.~4, no.~10, p. e7487, jan 2009.

\bibitem{siwiak2009catch}
D.~Siwiak, J.~Berger, and Y.~Yang, ``Catch your breath-musical biofeedback for
  breathing regulation,'' in \emph{Audio Eng. Society Conv.}, 2009.

\bibitem{yu2018unwind}
B.~Yu, M.~Funk, J.~Hu, and L.~Feijs, ``Unwind: a musical biofeedback for
  relaxation assistance,'' \emph{Behav. \& Info. Tech.}, vol.~37, no.~8, 2018.

\bibitem{vidyarthi2014interactively}
J.~Vidyarthi and B.~E. Riecke, ``Interactively mediating experiences of
  mindfulness meditation,'' \emph{International J. of Human-Computer Studies},
  vol.~72, no.~8, pp. 674--688, 2014.

\bibitem{robb2000music}
S.~L. Robb, ``Music assisted progressive muscle relaxation, progressive muscle
  relaxation, music listening, and silence: A comparison of relaxation
  techniques,'' \emph{J. of Music Therapy}, vol.~37, no.~1, pp. 2--21, 2000.

\bibitem{bernardi2006cardiovascular}
L.~Bernardi, C.~Porta, and P.~Sleight, ``Cardiovascular, cerebrovascular, \&
  respiratory changes induced by different types of music in musicians \&
  non-musicians: the importance of silence,'' \emph{Heart}, vol.~92, no.~4,
  2006.

\bibitem{bhandari2015music}
R.~Bhandari, A.~Parnandi, E.~Shipp, B.~Ahmed, and R.~Gutierrez-Osuna,
  ``Music-based respiratory biofeedback in visually-demanding tasks.'' in
  \emph{NIME}, 2015, pp. 78--82.

\bibitem{moraveji2011peripheral}
N.~Moraveji, B.~Olson, T.~Nguyen, M.~Saadat, Y.~Khalighi, R.~Pea, and J.~Heer,
  ``Peripheral paced respiration: influencing user physiology during
  information work,'' in \emph{ACM symp. on UI sw. \& tech.}, 2011.

\bibitem{robertson1982central}
R.~J. Robertson, ``Central signals of perceived exertion during dynamic
  exercise.'' \emph{Medicine \& Science in Sports \& Exercise}, vol.~14, no.~5,
  1982.

\bibitem{charmandari2005endocrinology}
E.~Charmandari, C.~Tsigos, and G.~Chrousos, ``Endocrinology of the stress
  response,'' \emph{Annu. Rev. Physiol.}, vol.~67, pp. 259--284, 2005.

\bibitem{bloch1991specific}
S.~Bloch, M.~Lemeignan, and N.~Aguilera-T, ``Specific respiratory patterns
  distinguish among human basic emotions,'' \emph{International J. of
  Psychophysiology}, vol.~11, no.~2, pp. 141--154, 1991.

\bibitem{brown2005sudarshana}
R.~P. Brown and P.~L. Gerbarg, ``Sudarshan kriya yogic breathing in the
  treatment of stress, anxiety, and depression: part i—neurophysiologic
  model,'' \emph{J. of Alternative \& Complementary Medicine}, vol.~11, no.~1,
  pp. 189--201, 2005.

\bibitem{brown2005sudarshanb}
------, ``Sudarshan kriya yogic breathing in the treatment of stress, anxiety,
  and depression: part ii—clinical applications and guidelines,'' \emph{J. of
  Alternative \& Complementary Medicine}, vol.~11, no.~4, 2005.

\bibitem{soni2015effect}
S.~Soni, L.~N. Joshi, and A.~Datta, ``Effect of controlled deep breathing on
  psychomotor \& higher mental functions in normal individuals.'' 2015.

\bibitem{philippot2002respiratory}
P.~Philippot, G.~Chapelle, and S.~Blairy, ``Respiratory feedback in the
  generation of emotion,'' \emph{Cognition \& Emotion}, vol.~16, no.~5, 2002.

\bibitem{bryant2015yoga}
E.~F. Bryant, \emph{The yoga sutras of Patanjali: A new edition, translation,
  and commentary}.\hskip 1em plus 0.5em minus 0.4em\relax North Point Press,
  2015.

\bibitem{kabat2003mindfulness}
J.~Kabat-Zinn, ``Mindfulness-based interventions in context: past, present, and
  future,'' \emph{Clinical psychology: Science and practice}, vol.~10, no.~2,
  pp. 144--156, 2003.

\bibitem{adams2015mindless}
A.~T. Adams, J.~Costa, M.~F. Jung, and T.~Choudhury, ``Mindless computing:
  designing technologies to subtly influence behavior,'' in \emph{Proceedings
  of the 2015 ACM International Joint Conference on Pervasive and Ubiquitous
  Computing}.\hskip 1em plus 0.5em minus 0.4em\relax ACM, 2015, pp. 719--730.

\bibitem{kuikkaniemi2010influence}
K.~Kuikkaniemi, T.~Laitinen, M.~Turpeinen, T.~Saari, I.~Kosunen, and N.~Ravaja,
  ``The influence of implicit and explicit biofeedback in first-person shooter
  games,'' in \emph{CHI}.\hskip 1em plus 0.5em minus 0.4em\relax ACM, 2010, pp.
  859--868.

\bibitem{ghandeharioun2017brightbeat}
A.~Ghandeharioun and R.~Picard, ``Brightbeat: Effortlessly influencing
  breathing for cultivating calmness and focus,'' in \emph{Proceedings of the
  2017 CHI Conference Extended Abstracts on Human Factors in Computing
  Systems}.\hskip 1em plus 0.5em minus 0.4em\relax ACM, 2017, pp. 1624--1631.

\bibitem{puckette1996pure}
M.~Puckette \emph{et~al.}, ``Pure data: another integrated computer music
  environment,'' \emph{Proceedings of the second intercollege computer music
  concerts}, pp. 37--41, 1996.

\bibitem{lehrer2000resonant}
P.~M. Lehrer, E.~Vaschillo, and B.~Vaschillo, ``Resonant frequency biofeedback
  training to increase cardiac variability: Rationale and manual for
  training,'' \emph{Applied psychophysiology and biofeedback}, vol.~25, no.~3,
  pp. 177--191, 2000.

\bibitem{Tecce1972}
J.~J. Tecce, ``{Contingent negative variation (CNV) and psychological processes
  in man},'' \emph{Psychological Bul.}, vol.~77, no.~2, pp. 73--108, 1972.

\bibitem{kothe2014lab}
C.~Kothe, ``Lab streaming layer (lsl),'' accessed: 2017-10-01.

\bibitem{delorme2004eeglab}
A.~Delorme and S.~Makeig, ``Eeglab: an open source toolbox for analysis of
  single-trial eeg dynamics including independent component analysis,''
  \emph{J. of neuroscience methods}, vol. 134, no.~1, pp. 9--21, 2004.

\bibitem{Funderud2012}
I.~Funderud, M.~Lindgren, M.~L{\o}vstad, T.~Endestad, B.~Voytek, R.~T. Knight,
  and A.~K. Solbakk, ``{Differential Go/NoGo Activity in Both Contingent
  Negative Variation and Spectral Power},'' \emph{PLoS ONE}, vol.~7, no.~10,
  2012.

\bibitem{pan1985real}
J.~Pan and W.~J. Tompkins, ``A real-time qrs detection algorithm,'' \emph{IEEE
  Trans. Biomed. Eng}, vol.~32, no.~3, pp. 230--236, 1985.

\bibitem{hrvpackage}
``Python package for heart rate variability analysis,''
  \url{https://github.com/rhenanbartels/hrv}, accessed: 2019-04-01.

\bibitem{frigge1989some}
M.~Frigge, D.~C. Hoaglin, and B.~Iglewicz, ``Some implementations of the
  boxplot,'' \emph{The American Statistician}, vol.~43, no.~1, pp. 50--54,
  1989.

\bibitem{grossman2001breathing}
E.~Grossman, A.~Grossman, M.~Schein, R.~Zimlichman, and B.~Gavish,
  ``Breathing-control lowers blood pressure,'' \emph{J. of human hypertension},
  vol.~15, no.~4, p. 263, 2001.

\bibitem{vlemincx2012sigh}
E.~Vlemincx, I.~Van~Diest, and O.~Van~den Bergh, ``A sigh following sustained
  attention and mental stress: effects on respiratory variability,''
  \emph{Physiology \& behavior}, vol. 107, no.~1, pp. 1--6, 2012.

\bibitem{vlemincx2011sigh}
E.~Vlemincx, J.~Taelman, S.~De~Peuter, I.~Van~Diest, and O.~Van Den~Bergh,
  ``Sigh rate and respiratory variability during mental load and sustained
  attention,'' \emph{Psychophysiology}, vol.~48, no.~1, 2011.

\bibitem{vlemincx2010take}
E.~Vlemincx, J.~Taelman, I.~Van~Diest, and O.~Van~den Bergh, ``Take a deep
  breath: the relief effect of spontaneous and instructed sighs,''
  \emph{Physiology \& behavior}, vol. 101, no.~1, pp. 67--73, 2010.

\bibitem{wysocki2006reduced}
M.~Wysocki, C.~Cracco, A.~Teixeira, A.~Mercat, J.-L. Diehl, Y.~Lefort, J.-P.
  Derenne, and T.~Similowski, ``Reduced breathing variability as a predictor of
  unsuccessful patient separation from mechanical ventilation,'' \emph{Critical
  care medicine}, vol.~34, no.~8, pp. 2076--2083, 2006.

\bibitem{baudin2014impact}
F.~Baudin, H.-T. Wu, A.~Bordessoule, J.~Beck, P.~Jouvet, M.~G. Frasch, and
  G.~Emeriaud, ``Impact of ventilatory modes on the breathing variability in
  mechanically ventilated infants,'' \emph{Frontiers in pediatrics}, vol.~2, p.
  132, 2014.

\bibitem{boucsein2012electrodermal}
W.~Boucsein, \emph{Electrodermal activity}.\hskip 1em plus 0.5em minus
  0.4em\relax Springer Sci. \& Bus. Media, 2012.

\bibitem{LIM1996151}
C.~Lim, R.~Barry, E.~Gordon, A.~Sawant, C.~Rennie, and C.~Yiannikas, ``The
  relationship between quantified eeg and skin conductance level,''
  \emph{International J. of Psychophysiology}, vol.~21, no.~2, pp. 151 -- 162,
  1996.

\bibitem{higuchi1997effects}
S.~Higuchi, S.~Watanuki, and A.~Yasukouchi, ``Effects of reduction in arousal
  level caused by long-lasting task on cnv,'' \emph{Applied Human Science},
  vol.~16, no.~1, pp. 29--34, 1997.

\bibitem{nagai2004influence}
Y.~Nagai, L.~H. Goldstein, H.~D. Critchley, and P.~B. Fenwick, ``Influence of
  sympathetic autonomic arousal on cortical arousal: implications for a
  therapeutic behavioural intervention in epilepsy,'' \emph{Epilepsy research},
  vol.~58, no. 2-3, pp. 185--193, 2004.

\bibitem{aysin2006effect}
B.~Aysin and E.~Aysin, ``Effect of respiration in heart rate variability (hrv)
  analysis,'' in \emph{EMBC}.\hskip 1em plus 0.5em minus 0.4em\relax IEEE,
  2006, pp. 1776--1779.

\bibitem{picard2016multiple}
R.~W. Picard, S.~Fedor, and Y.~Ayzenberg, ``Multiple arousal theory and
  daily-life electrodermal activity asymmetry,'' \emph{Emotion Rev.}, vol.~8,
  no.~1, pp. 62--75, 2016.

\end{thebibliography}

\end{document}